# Coefficient of performance at maximum cooling power of a simplified quantum dot refrigerator model with resistance

Xiaolong Liu[a,b,*], Xiaoguang Luo[c,*], Chengyun Hong[b], Zhen Bao[b], Yong Ding[a,b], Jianxi Yao[a,b], and Songyuan Dai[a,b]

[a] *Beijing Key Laboratory of Novel Thin-Film Solar Cells & Beijing Key Laboratory of Energy Safety and Clean Utilization, North China Electric Power University, Beijing 102206, China*
[b] *Renewable Energy School, North China Electric Power University, Beijing 102206, China*
[c] *Xi'an Institute of Flexible Electronics & MIIT Key Laboratory of Flexible Electronics & Shaanxi Key Laboratory of Flexible Electronics & Xi'an Key Laboratory of Flexible Electronics, Northwestern Polytechnical University, Xi'an 710072, China*

**ABSTRACT**

A simplified analytical model of single-level quantum dot (QD) refrigerator was studied without considering the electron spin and Coulomb interaction. Based on the ballistic transport of electrons between two reservoirs across the QD, the Joule heat of the system was assumed to be generated from the Ohmic contacts between the QD and reservoirs. By using the transition rate equation, the performance of the QD refrigerator was studied with respect to the electron transmission probability and the partition ratio (i.e., the fraction of Joule heat generated in the system that releases into the cold reservoir). The analytical expression of the maximum coefficient of performance was obtained under the exoreversible working condition. The Carnot-bound-dependent coefficient of performance at maximum cooling power of the QD system was also demonstrated numerically. The results of this work may provide some guidance for the design of mesoscopic refrigerators.

## 1. Introduction

It is a promising way adopting thermoelectric effect to realize effective conversion between thermal and electric energies at mesoscopic size. Many breakthroughs successively refreshed the state of art of thermoelectric applications and even draw remarkable research interest in this field. For examples, a new record high figure of merit $ZT\sim2.8$ has been recently found in Br-doped SnSe [1], the thermodynamic limits were approached by an InP/InAs/InP quantum dot heat engine [2], and especially, some mesoscopic thermoelectric devices based on molecular junctions were successfully designed for power generation [3] or Peltier cooling [4,5]. Therein, thermoelectric refrigerators have been widely used in industry and our daily life. Driven by the input electric power, the thermoelectric refrigerator absorbs heat from the cold reservoir (with temperature $T_c$) and releases heat to the hot reservoir (with temperature $T_h$). From the second law of thermodynamics, the coefficient of performance (COP) of any refrigerator cannot exceed the Carnot value, $\epsilon_C = T_c/(T_h - T_c) = \tau/(1-\tau)$ with $\tau = T_c/T_h$, which is the maximum COP for an infinite thermodynamic cycle duration, and the refrigerator is then at the reversible working state with the cooling power consequently vanishes [6]. The thermoelectric refrigerator is not an exception, the maximum COP tends to Carnot value at the reversible working state, without producing any cooling power [7,8].

An effective refrigerator should produce cooling power, thus it is necessary to find the COP at maximum cooling power (CMCP). However, the general form of $\epsilon_C$-dependent CMCP of most refrigerators does not exist without considering the system dissipation. Fortunately, by maximizing the product of COP and cooling power (i.e. the so-called $\chi$-criterion) [9,10], $\epsilon_C$-dependent COP at maximum $\chi$ can be obtained for some endoreversible refrigerators. For example, it was found from the Carnot-like refrigerators by the finite-time technique [11] that, the COP at maximum $\chi$ is bounded by an elegant expression of $(\sqrt{9+8\epsilon_C}-3)/2$. This boundary, seems to be universal, is larger than all the obtained COPs at maximum $\chi$ in Fermi-Dirac system, Maxwell-Boltzmann system, and Bose-Einstein system [12-14]. Therein, the dissipation is actually considered, which comes from the coupling between the working system and the external heat reservoirs.

*E-mail: xl.liu@ncepu.edu.cn (XLiu); iamxgluo@nwpu.edu.cn (XLuo)

Realizing $\chi$-criterion was still not the way to obtain the true CMCP, Apertet *et al.* [15] studied a generalized thermoelectric refrigerator with Ohmic resistance (a so-called exoreversible refrigerator) and found the $\epsilon_C$-dependent CMCP $\epsilon_A = \epsilon_C/(2 + \epsilon_C/\lambda)$, where $\lambda$ was the partition ratio of the internal dissipations (Joule heat from Ohmic resistance) releasing into the cold reservoir compared to the total Joule heat generated in the system. This expression indicates that the CMCP is an increasing function of $\lambda$. It should be noted that the thermoelectric refrigerator in reference [15] is a macroscopic model, in which the dimension of the conductor is much larger than the mean free path of electrons. Therefore, the Ohmic resistance $R = \rho l/s$ can be treated as a constant, where $\rho$ is the resistivity, $l$ and $s$ are, respectively, the length and cross-sectional area of the conductor. When the thermoelectric refrigerator goes down to mesoscopic, the dissipation behavior will be distinct from Ohm's law.

In a mesoscopic system, electrons could transport ballistically if the conductor length is smaller than the electron mean free path. As a result, the resistance from electron-electron and electron-phonon interactions can be ignored along the electron transport path, and basically the Joule heat comes from the Ohmic contacts between the working system and the reservoirs [16]. In this paper, a mesoscopic refrigerator based on a single-level quantum dot was studied, the CMCP then could be found by taking the Ohmic contacts into account. The organization of this paper is as follows. In the second section, we illustrated that the CMCP cannot be obtained from a general single-level quantum dot (QD) refrigerator without resistance. In the third section, we discussed two different kinds of resistance in our mesoscopic system and the origin of the Joule heat in the QD refrigerator system. In the fourth section, we investigated the impact of the electron transmission probability $t$ and partition ratio $\lambda$ on the CMCP. Finally, the conclusions were summarized at the end of this paper.

## 2. Quantum dot refrigerator

As the material size goes down to nanoscale, quantum confinement effect plays an important role in the electronic properties [17-20], reflected from the fact that the energy bands are divided into discrete energy levels. With the nanometer spatial size, there are only several or even one free electron in a QD owing to the strong Coulomb interaction [21]. In this work, a single-level QD was embedded between a cold and a hot reservoir (or electrode) with different temperature ($T_c < T_h$) and chemical potential ($\mu_c > \mu_h$), as seen in Fig. 1. Electrons can be exchanged between two reservoirs due to the chemical potential gradient and temperature gradient, thus the electron flux is generated together with the coupled heat flux, which can be manipulated by controlling the transmission behavior of electrons inside the QD system.

There is only none (state 0) or one (state 1) free electron in the single-level QD under the assumption of infinitely strong Coulomb interaction, with the energy level $\varepsilon$ locating in the "Fermi window" between two reservoirs, i.e., an energy region of several $k_B T$ around the reservoir chemical potential. Without considering electron spin, the Hamiltonian of the system can be expressed as [19,20]

$$H = \varepsilon \hat{a}^\dagger \hat{a} + \sum_{iv} \varepsilon_i \hat{r}_{iv}^\dagger \hat{r}_{iv} + \sum_{iv} c_{iv}(\hat{a}^\dagger \hat{r}_{iv} + \hat{a} \hat{r}_{iv}^\dagger) \quad (1)$$

where the three terms describe the QD, the reservoirs, and their interaction orderly. $\hat{a}^\dagger(\hat{a})$ is the creation (annihilation) operator on the energy level $\varepsilon$, $\hat{r}_{iv}^\dagger(\hat{r}_{iv})$ is the creation (annihilation) operator on the level $i$ with energy $\varepsilon_i$ of reservoir $\nu$ ($\nu = c, h$, $c$ and $h$ represent "cold" and "hot", respectively). $c_{iv}$ is the coupling coefficient of the interaction between the QD and the reservoirs, denoting the electron transition between the energy level $\varepsilon$ and $\varepsilon_i$.

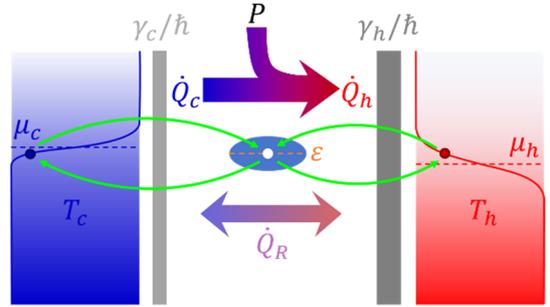

**Fig. 1.** Sketch of a single-level (with energy $\varepsilon$) QD refrigerator with resistance. The temperatures and chemical potentials of cold and hot reservoirs fulfill $T_c < T_h$ and $\mu_c > \mu_h$, respectively. $\gamma_{c/h}/\hbar$ is the bare transition rate of electrons between the cold/hot reservoir and the QD. Driven by the input power $P$, electrons absorb heat flux $\dot{Q}_c$ from the cold reservoir and then release heat flux $\dot{Q}_h$ into the hot reservoir, the Joule heat flux $\dot{Q}_R$ generated during this process finally flows into the reservoirs.

Generally, the bare transition rate between the QD and a reservoir can be described by Lorentzian resonance $\Lambda_{iv} = \gamma_v/\hbar\{1 + [(\varepsilon - \varepsilon_{iv})/W_v]^2\}$, where $\hbar$ is the reduced Planck constant, $\varepsilon_{iv}$ is the energy of level $i$ in reservoir $\nu$ ($\nu = c, h$, represents "cold" or "hot", respectively), and $W_v$ is the bandwidth which is related to the coupling between the QD and reservoirs. It should be noted that the broader electron transmission spectrum will lower the maximum COP and the CMCP [12,14]. To obtain the upper bound of COP and CMCP, the weak coupling was considered in this paper, i.e., $W_v \to 0$ and $\Lambda_{iv} = \gamma_v/\hbar \times \delta(\varepsilon - \varepsilon_{iv})$. Thus, only the electrons with energy $\varepsilon$ contribute to the electron current, and the bare transition rate $\Lambda_v = \gamma_v/\hbar$ is strongly dependent on the contact property. The occupation probability of the reservoirs obeys the Fermi-Dirac distribution $f_v = 1/(e^{x_v} + 1)$, where $x_v$ is the dimensional scaled energy expressed as $(\varepsilon - \mu_v)/k_B T_v$, and $k_B$ is the Boltzmann constant. Subsequently, the transition rate of QD from state 0 to state 1 can be counted by $\Gamma_{10} = \sum_{v=c,h} \Gamma_{dv} = \sum_{v=c,h} \Lambda_v f_v$, and from state 1 to state 0 as $\Gamma_{01} = \sum_{v=c,h} \Gamma_{vd} = \sum_{v=c,h} \Lambda_v (1-f_v)$, where $\Gamma_{dv}$ and $\Gamma_{vd}$ denote the transition rate between QD ($d$) and reservoir ($v$). Finally, the probability of empty/filled state of the QD, $p_{0/1}$ can be determined from the rate equation [22-24]

$$\begin{pmatrix} \dot{p}_0 \\ \dot{p}_1 \end{pmatrix} = \begin{pmatrix} -\Gamma_{10} & \Gamma_{01} \\ \Gamma_{10} & -\Gamma_{01} \end{pmatrix} \begin{pmatrix} p_0 \\ p_1 \end{pmatrix} \quad (2)$$

The probabilities in the steady state tend to be constant with respect to time and then $\Gamma_{10} p_0 - \Gamma_{01} p_1 = 0$. Combined with $p_0 + p_1 = 1$, $p_1$ is derived as $(\gamma_c f_c + \gamma_h f_h)/(\gamma_c + \gamma_h)$.

Furthermore, the electron current flowing out of the reservoir can be calculated from the transmission between one reservoir and the QD, expressed as $n_v = \Gamma_{dv} p_0 - \Gamma_{vd} p_1$ with $n_c + n_h = 0$. Once setting the direction from cold to hot reservoir to be positive, the electric current of the system $I = e n_c$ is written as

$$I = \frac{e}{\hbar} \frac{\gamma_c \gamma_h}{\gamma_c + \gamma_h} (f_c - f_h) \quad (3)$$

where $e = -1.602 \times 10^{-19}$ C is the elementary charge.

It is known that when an electron with energy $\varepsilon$ arrives/leaves a reservoir with chemical potential $\mu_v$, the reservoir will absorb/release an average amount of heat $\varepsilon - \mu_v$. Then the heat flux flowing out/in of the cold/hot reservoir is expressed as:

$$\begin{aligned} \dot{Q}_{c/h} &= \frac{1}{\hbar} \frac{\gamma_c \gamma_h}{\gamma_c + \gamma_h} (\varepsilon - \mu_{c/h})(f_c - f_h) \\ &= \frac{1}{\hbar} \frac{\gamma_c \gamma_h}{\gamma_c + \gamma_h} k_B T_{c/h} x_{c/h} (f_c - f_h) \end{aligned} \quad (4)$$

$\dot{Q}_c$ here is the cooling power of the refrigerator, and the corresponding input power is

$$P = (\dot{Q}_h - \dot{Q}_c) = \frac{1}{\hbar} \frac{\gamma_c \gamma_h}{\gamma_c + \gamma_h} (k_B T_h x_h - k_B T_c x_c)(f_c - f_h) \quad (5)$$

Considering that to generate cooling power by inputting power in other forms (electric power as an example) is the fundamental working mechanism of a refrigerator, it is straightforward that the cooling power and the input power should be positive simultaneously, i.e., $\dot{Q}_c \geq 0$ and $P \geq 0$, which result in $f_c \geq f_h$ and $x_h \geq x_c > 0$ since that $k_B T_h x_h - k_B T_c x_c = \mu_c - \mu_h > 0$ in our model. The refrigerator stops cooling when all equality signs hold in these relations, and the system tends to be at equilibrium state.

Subsequently, the COP of the refrigerator $\epsilon = \dot{Q}_c/P$ is expressed as

$$\epsilon = \frac{x_c T_c}{x_h T_h - x_c T_c} = \frac{\zeta}{1 + 1/\epsilon_C - \zeta} \quad (6)$$

where $0 < \zeta \equiv x_c/x_h \leq 1$, then it is straightforward to find that the maximum COP is the Carnot value $\epsilon_C$ when $\zeta = 1$. The entropy production rate of the system is $\dot{S} = -\dot{Q}_c/T_c + \dot{Q}_h/T_h = k_B \gamma_c \gamma_h / [\hbar(\gamma_c + \gamma_h)] \times (x_h - x_c)(f_c - f_h)$. It can be easily found $\dot{S} \geq 0$ because $f_v$ is a decreasing function of $x_v$. Approaching the working state with $\dot{S} = 0$, namely the reversible process, the COP of the refrigerator tends to be the Carnot value, and the refrigerator is reversible. However, all the fluxes vanish to zero, as well as the cooling power. To balance the cooling power and the COP, the CMCP instead was investigated in this paper as a more practical parameter.

From the cooling power expression of Eq. (4), $x_c \neq 0$ as $\dot{Q}_c \neq 0$ in the cooling process, resulting in $\partial \dot{Q}_c / \partial x_h = x_c e^{x_h}/(e^{x_h}+1)^2 \neq 0$. In other words, the $\epsilon_C$-dependent CMCP cannot be found without considering the internal dissipation.

Taking the Ohmic resistance into account, Apertet et al. [15] obtained that the $\epsilon_C$-dependent CMCP could be expressed as $\epsilon_A = \epsilon_C/(2 + \epsilon_C/\lambda)$ for a macroscopic thermoelectric refrigerator model, where $\lambda$ is the partition ratio. We next show that the CMCP can also be found for the mesoscopic single-level QD refrigerator with the consideration of the Ohmic resistance.

## 3. QD Refrigerators with resistance

In a mesoscopic system, the resistance is not constant anymore and should be calculated by considering the electrical conductivity behavior. The electric current of the QD refrigerator is rewritten as

$$I = \frac{e}{h} \frac{\Gamma}{2\pi} t(f_c - f_h) \quad (7)$$

where $t = \gamma_c \gamma_h / [\Gamma(\gamma_c + \gamma_h)]$ is defined as the transmission probability. The conductance is then expressed as

$$G = \frac{I}{(\mu_c - \mu_h)/e} = \frac{e^2}{h} \frac{\Gamma}{2\pi} \frac{f_c - f_h}{\mu_c - \mu_h} t \quad (8)$$

At low temperature limit, the conductance becomes $G = (e^2/h) \times \{\Gamma/[2\pi(\mu_c - \mu_h)]\} t$ in the "Fermi window", which corresponds to the two-terminal Landauer formula $G = (2e^2/h) \times Mt$, where $M \to \Gamma/[2\pi(\mu_c - \mu_h)]$ tells the number of modes above cut-off, and the factor "2" in Landauer formula denotes the spin up or down of an electron [25], which is constant and is not counted in our model. The resistance of the QD system then can be expressed as $G^{-1}$, though is not the actual ohmic resistance. It is known that ohmic resistance rises along with the electrons scattering and then the resistance of the QD refrigerator is written as [26]

$$G^{-1} = \frac{h}{e^2} \frac{2\pi}{\Gamma} \frac{\mu_c - \mu_h}{f_c - f_h} + \frac{h}{e^2} \frac{2\pi}{\Gamma} \frac{\mu_c - \mu_h}{f_c - f_h} \frac{r}{t} \quad (9)$$

where $r = 1 - t$ is the reflection coefficient. Ignoring other forms of reflection than scattering, the first term in Eq. (9) is the contact resistance $G_C^{-1}$ and the second term is the scattering resistance $G_S^{-1}$[26]. Generally, the typical mean free path of electrons in a macroscopic conductor is about tens or hundreds of nanometers [27], and the relevant transmission probability of the electron $t \to 0$, which results in $G_S^{-1} \gg G_C^{-1}$. Then the contact resistance can be ignored, and the resistance is regarded as the actual ohmic resistance. While for a mesoscopic conductor, the contact resistance $G_C^{-1}$ must be considered since the transmission probability $t$ is comparable to the reflection coefficient $r$ of electrons. Along with the fact that the contact resistance $G_C^{-1}$ is related to the ballistic current and does not dissipate heat [28], therefore, only the scattering resistance $G_S^{-1}$ contributes to the Joule heat. At low bias, the Joule heat goes quadratically with bias [29], and the bias here is the voltage drop $U_S = IG_S^{-1}$ due to the scattering resistance. The total Joule heat then can be calculated as

$$\dot{Q}_R = U_S^2/G_S^{-1} = \frac{1}{h}\frac{\Gamma}{2\pi}t(1-t)(\mu_c - \mu_h)(f_c - f_h) \quad (10)$$

It should be noted that this expression is under the condition of $\zeta \leq 1$. For the case of $\zeta > 1$, where $f_c < f_h$, the total Joule heat should be expressed as $\dot{Q}_R = \frac{1}{h}\frac{\Gamma}{2\pi}t(1-t)(\mu_c - \mu_h)(f_h - f_c)$. Moreover, it is found that the Joule heat increases initially as $t$ increases to 1/2, and then decreases as $t$ increases to 1, because the electron current is also related to $t$. In the QD refrigerator, the Joule heat is generated during the electron transition process between the QD and two different reservoirs where the scattering occurs. However, it is adopted that Joule heat associated with resistance $G_S^{-1}$ goes locally into the overall electron system [26]. We assume that certain amount of Joule heat $\lambda \dot{Q}_R$ flows into the cold reservoir, and the residual Joule heat $(1-\lambda)\dot{Q}_R$ flows into the hot reservoir, with the partition ratio $0 < \lambda < 1$. For a typical thermoelectric system [30, 31], $\lambda = 1/2$, while here we treat it as a variable parameter to find its dependence on the CMCP. Subsequently, from Eqs. (4) and (10), the cooling power becomes $\dot{Q}_{CP} = \dot{Q}_c - \lambda \dot{Q}_R$, i.e.,

$$\dot{Q}_{CP} = A_c t(f_c - f_h)\{x_c - \lambda(1-t) \times [(1+1/\epsilon_C)x_h - x_c]\} \quad (11)$$

where $A_c = \frac{k_B T_c}{h}\frac{\Gamma}{2\pi}$, and the power input is

$$P = A_c t(f_c - f_h)(2-t)[(1+1/\epsilon_C)x_h - x_c] \quad (12)$$

From $\dot{Q}_{CP} \geq 0$, $P \geq 0$ and $\mu_c > \mu_h$, the results of $f_c \geq f_h$ and $1 \geq \zeta \geq (1+1/\epsilon_C)/[1+1/\lambda(1-t)]$ can be found similarly as that without considering the resistance. The COP then is

$$\epsilon = \frac{\zeta - \lambda(1-t)(1+1/\epsilon_C - \zeta)}{(2-t)(1+1/\epsilon_C - \zeta)} \quad (13)$$

One can find that the maximum COP $\epsilon_{max} = [\epsilon_C - \lambda(1-t)]/(2-t)$ is smaller than the Carnot COP except when the transmission probability reaches to 1. The maximum COP determines the thermoelectric figure of merit of $ZT$ from the relationship $\epsilon_{max} = \epsilon_C(\sqrt{1+ZT} - 1/\tau)/(\sqrt{1+ZT} + 1)$ [32]. Then, the upper bound $ZT_{max} = [(3-t)/(1-t)]^2 - 1$ is obtained at $\epsilon_C \to \infty$. The $ZT_{max}$ only depends on the transmission probability $t$, other than the partition ratio $\lambda$. It is consistent with the original definition of $ZT = Gs^2T/k$, where $s$ is Seebeck coefficient and $k$ is the thermal conductivity due to electron and lattice [33]. The increased transmission probability $t$ benefits the conductivity $G$ (see Eq. (8)) and finally enhances $ZT$.

The entropy production rate of the refrigerator now is

$$\dot{S} = \frac{k_B}{h}\frac{\Gamma}{2\pi}t(f_c - f_h)\{(x_h - x_c) + \frac{\lambda+\epsilon_C}{1+\epsilon_C}(1-t) \times [(1+1/\epsilon_C)x_h - x_c]\} \quad (14)$$

in which $\dot{S} \geq 0$, and the equality sign holds if $x_h = x_c$ or $t = 0$. Ignoring the trivial situation of $t = 0$, the maximum COP $\epsilon_{max}$ cannot reach to Carnot value due to $t \neq 1$, even if the system tends to equilibrium state (i.e. $\dot{S} = 0$) and therefore the refrigerator is exoreversible.

## 4. COP at maximum cooling power

A first observation from Eq. (11) is that the cooling power could approach its maximum value as the transmission probability $t$ increases. From $\partial \dot{Q}_{CP}/\partial t = 0$, the optimum $t = 1/2 \times \{1 - \zeta/[\lambda(1+1/\epsilon_C - \zeta)]\}$ with respect to $\zeta = x_c/x_h$ is obtained. Then, the cooling power becomes $\dot{Q}_{CP} = -A_c x_h(f_c - f_h)[\lambda(1+1/\epsilon_C - \zeta) - \zeta]^2/[4\lambda(1+1/\epsilon_C - \zeta)]$. It has been mentioned previously that $f_c \geq f_h$ and $\zeta \leq 1$, thus the optimum cooling power $\dot{Q}_{CP} \leq 0$ here, and therefore the cooling power cannot be maximized with respect to $t$ in such way. We then calculated the cooling power and the relevant COP of the QD refrigerator with respect to the dimensionless scaled energies $x_{c/h}$ at the given partition ratio $\lambda = 1/2$ and transmission probability $t$, as shown in Fig. 2. It is found that $(1+1/\epsilon_C)/[1+1/\lambda(1-t)] \leq \zeta \leq 1$ in the cooling region, and the COP reaches the maximum value $\epsilon_{max}$ at $x_c = x_h$. Higher transmission probability $t$ affords larger cooling power and COP. When $t = 1$, i.e. the situation without resistance, it is found that no maximum cooling power can be obtained with respect to $x_{c/h}$. Once taking the resistance into account, as seen from Figs. 2(a) and 2(b), one can get the maximum cooling power by optimizing $x_{c/h}$, and then the CMCP can be calculated.

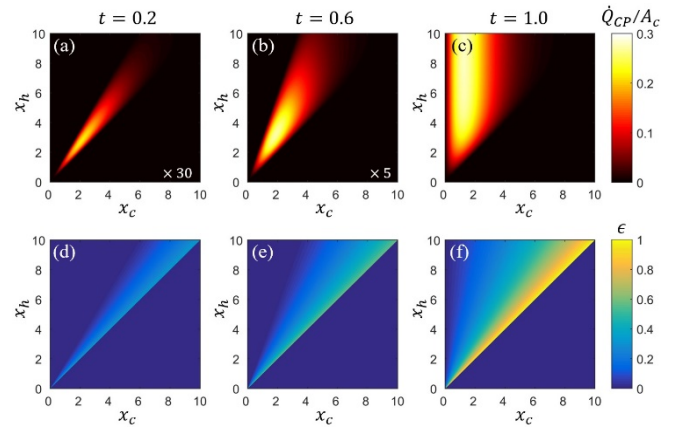

**Fig. 2.** The cooling power of the single-level QD refrigerator with electron transmission probability (a) $t = 0.2$, (b) $t = 0.6$, and (c) $t = 1.0$, after fixing $\lambda = 1/2$, and $\epsilon_C = 1$. (d)-(f) are the corresponding COPs.

Imposing $\partial \dot{Q}_{CP}/\partial x_c = \partial \dot{Q}_{CP}/\partial x_h = 0$, the optimized

value of $x_{c/h}$ at maximum cooling power are figured out by

$$(f_c - f_h)[1 + \lambda(1-t)] = e^{x_c}f_c^2\{x_c - \lambda(1-t)[(1+1/\epsilon_C)x_h - x_c]\} \quad (15a)$$

$$(f_c - f_h)\lambda(1-t)(1+1/\epsilon_C) = e^{x_h}f_h^2\{x_c - \lambda(1-t)[(1+1/\epsilon_C)x_h - x_c]\} \quad (15b)$$

from which it is found that the maximum cooling power is related to $\lambda(1-t)$ directly, and $0 < \lambda(1-t) < 1$. By solving the equation set above,

$$x_c = 2\ln\left[\frac{1}{\sqrt{\beta}}\cosh(x_h/2) + \sqrt{\frac{1}{\beta}\cosh^2(x_h/2) - 1}\right] \quad (16)$$

where $\beta = [1 + \lambda(1-t)]/[\lambda(1-t)(1+1/\epsilon_C)]$.

Substituting Eq. (16) into Eq. (15a) or (15b), a transcendental equation is obtained, and $x_h$ can be calculated numerically.

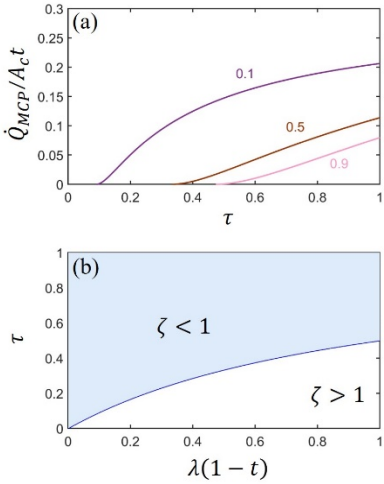

Fig. 3. (a) The maximum cooling power as a function of temperature ratio $\tau$ with $\lambda(1-t) = 0.1$, 0.5, and 0.9, respectively. (b) The optimized cooling region (light blue region) and cut off bound (blue solid line) with respect to $\tau$ and $\lambda(1-t)$.

According to Eqs. (11) and (15-16), we then calculated the cut-off bound of the maximum cooling power with respect to the term $\lambda(1-t)$ and the temperature ratio $\tau$ (or the Carnot COP $\epsilon_C = \tau/(1-\tau)$ equivalently). As seen from Fig. 3(a), maximum cooling power $\dot{Q}_{MCP}$ increases with $\tau$ at the fixed $t$, and the zero point of $\dot{Q}_{MCP}$ at the $\tau$ axis increases with $\lambda(1-t)$. At those $\dot{Q}_{MCP}$ zero points, where $f_c = f_h$ (or $\zeta = 1$) according to Eq. (11), no current and heat flux can be generated. In the region at the right side of zero points, we found that $\zeta < 1$, as well as the definite maximum cooling power. No solution under maximum cooling power can be obtained in those regions at the left side of zero points with the restricted condition of $\zeta \leq 1$, in other words, $\zeta$ should be large than 1 there and Eq. (11) is no longer valid for cooling power. Instead, the total Joule heat in the cooling power expression $\dot{Q}_{CP} = \dot{Q}_c - \lambda\dot{Q}_R$ should be replaced by $\dot{Q}_R = \frac{1}{h}\frac{\Gamma}{2\pi}t(1-t)(\mu_c - \mu_h)(f_h - f_c)$. Here $\dot{Q}_{CP} < 0$ under the condition of $\zeta > 1$ because of $\dot{Q}_c <$ 0 and $\dot{Q}_R > 0$. Therefore, the cold reservoir is heated in the regions by the left side of zero points, and then the refrigerator stops working. These zero points subsequently form a cut-off bound.

On the cut-off bound (the blue solid line in Fig.3(b)), the maximum cooling power vanishes (i.e., $\dot{Q}_{MCP} = 0$ and $\partial\dot{Q}_{CP}/\partial x_c = \partial\dot{Q}_{CP}/\partial x_c = 0$), resulting in $\epsilon_C = \lambda(1-t)$. Two regions associated with $\tau$ and $\lambda(1-t)$ are divided by the cut-off bound, including the maximum cooling power region (the light blue region where $\zeta < 1$) and the heating region (the white region where $\zeta > 1$). The maximum cooling power can be obtained by optimizing $x_{c/h}$ only in the region with $\epsilon_C \geq \lambda(1-t)$, i.e. the light blue region.

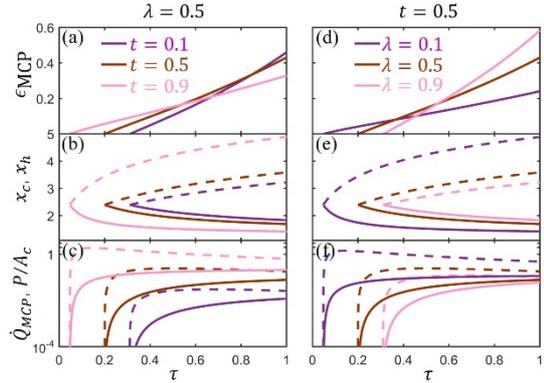

Fig. 4. For $\lambda = 0.5$, CMCPs (a), the corresponding dimensionless scaled energies (b), and the maximum cooling powers and power inputs (c) with respect to temperature ratio $\tau$ when $t = 0.1$, 0.5, and 0.9, respectively. For $t = 0.5$, CMCPs (d), the corresponding dimensionless scaled energies (e), and the maximum cooling powers and power inputs (f) with respect to temperature ratio $\tau$ when $\lambda = 0.1$, 0.5, and 0.9, respectively. (c) and (d) are semi-log plots. The solid lines and dashed lines are for $x_c$ and $x_h$ respectively in (b) and (e), and for maximum cooling power and the corresponding power inputs respectively in (c) and (f).

Unlike the cooling power, the COP depends not only on $\lambda(1-t)$, but also on both the partition ratio $\lambda$ and the transmission probability $t$, as shown in Eq. (13). At different $\lambda$ and $t$, the maximum cooling power can be calculated numerically from Eq. (14) as a function of the temperature ratio $\tau$ by optimizing $x_{c/h}$, then one can derive the power input and the CMCP.

At fixed $\lambda = 0.5$, Fig. 4(c) depicts the maximum cooling powers (solid lines) and the corresponding power inputs (dashed lines) as the function of $\tau$ when $t = 0.1$, $t = 0.5$, and $t = 0.9$, respectively, from which we get the corresponding CMCPs, as shown in Fig. 4(a), with the corresponding $x_c$ (solid lines) and $x_h$ (dashed lines) in Fig. 4(b). It is found that the maximum cooling powers and CMCPs increase as $\tau$ increases, while the corresponding power inputs increase initially and then decrease to finite values. Similar phenomena are found for $\lambda = 0.1$, $\lambda = 0.5$, and $\lambda = 0.9$ with fixed $t = 0.5$, as shown in Figs. 4(d), 4(e), and 4(f). It also shows that the performance of the single-level QD refrigerator at maximum cooling power is influenced dramatically by the cut-off bound $\epsilon_C = \lambda(1-t)$ for relatively small $\tau$. Larger $t$ or smaller $\lambda$ seems to gain the maximum cooling power and CMCP more effectively at small $\tau$, where more electrons are transported from the cold reservoir to hot one and less Joule heat flows into the cold reservoir. While at certain large $\tau$ close to 1 ($\epsilon_C \to \infty$), CMCP increases when $t$ decreases or when $\lambda$ increases though with the maximum cooling power always decreases, in other words, less power input is needed for a certain amount of maximum cooling power at smaller $t$ and larger $\lambda$.

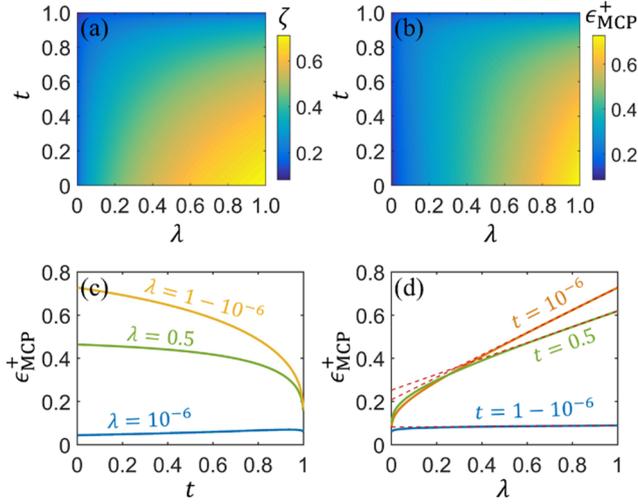

**Fig. 5.** When Carnot COP $\epsilon_C \to \infty$, (a) the ratio of corresponding dimensionless scaled energies and (b) CMCP with respect to $t$ and $\lambda$. (c) CMCP as a function of $t$ when $\lambda = 10^{-6}, 0.5$, and $1 - 10^{-6}$. (d) CMCP as a function of $\lambda$ when $t = 10^{-6}, 0.5$, and $1 - 10^{-6}$, the dashed lines are the fitted lines $\epsilon_{MCP}^+ = a + b\lambda$ by the data in the range of $\lambda \in (0.8, 1.0)$, with the corresponding parameters of $(0.20855, 0.51903)$, $(0.25307, 0.36625)$, and $(0.08253, 0.00685)$ for $(a, b)$, respectively.

A common conclusion is that the CMCP is an increasing function of the temperature ratio $\tau$ (or the Carnot COP $\epsilon_C$ equivalently). Therefore, the upper bound of CMCP can be obtained at $\epsilon_C \to \infty$, as

$$\epsilon_{MCP}^+ = \frac{\zeta}{1-\zeta}\frac{1}{2-t} - \lambda\frac{1-t}{2-t} \quad (17)$$

where $\zeta$ can be calculated from Eq. (15) when $\epsilon_C \to \infty$. The expression here is intuitively different from the CMCP upper bound in the macroscopic thermoelectric refrigerator model [11] with $\epsilon_A^+ = \lambda$. In the mesoscopic QD refrigerator, $\zeta$ is also the function of $\lambda$ and $t$, and therefore the CMCP upper bound $\epsilon_{MCP}^+$ varies with both $\lambda$ and $t$ (see Figs. 5a and 5b). At $(0, 1)$ and $(1, 0)$ of $(\lambda, t)$, $\zeta$ tends to be the minimum value of 0.034076 and the maximum value of 0.71056, respectively. The corresponding $\epsilon_{MCP}^+$ approaches the minimum value 0.035278 and the maximum value 0.72748.

To obtain more details, we firstly plotted the curves of CMCP upper bound with respect to the transmission probability $t$ after fixing the partition ratio $\lambda = 10^{-6}, 0.5$, and $1 - 10^{-6}$, respectively, as shown in Fig. 5(c). At very small $\lambda \sim 10^{-6}$, $\epsilon_{MCP}^+$ is also very small, and a peek can be found as $t$ increases. Larger $\epsilon_{MCP}^+$ can be obtained at larger $\lambda$. When $\lambda$ is larger enough (e.g., $\lambda = 0.5$, and $1 - 10^{-6}$), $\epsilon_{MCP}^+$ will decrease monotonously with respect to $t$. Then, we investigated the relationship between the CMCP upper bound and the partition ratio $\lambda$ at $t = 10^{-6}, 0.5$, and $1 - 10^{-6}$, respectively, as shown in Fig. 5(d). The difference is that larger $\epsilon_{MCP}^+$ is obtained more easily for smaller $t$, and the $\epsilon_{MCP}^+$ increases monotonously as $\lambda$ increases for any given $t$. In addition, $\epsilon_{MCP}^+$ tends to be sublinear as $\lambda$ increases. The numeral data in the range $\lambda \in (0.8, 1.0)$ is linearly fitted with the form $\epsilon_{MCP}^+ = a + b\lambda$. It is always found that $a \neq 0$ and $b > 0$. In other words, absorbing more Joule heat from the cold reservoir (larger $\lambda$) corresponds to larger CMCP, which is partially consistent with the case in the macroscopic thermoelectric refrigerator [15].

## 5. Conclusions

A mesoscopic refrigerator was established, in which a single-level QD was embedded between the cold and the hot reservoirs. Electrons were assumed to transport ballistically between two reservoirs. The Joule heat then can only be generated at the Ohmic contacts between the QD and reservoirs, stemming from the electron scattering, and the Ohmic resistance of the system was figured out from the reflection coefficient of electrons after ignoring other reflection forms other than scattering. Besides, the Joule heat flows into the reservoirs along with the electron transport, thus the QD refrigerator performance depends both on the electron transmission probability $t$ and the partition ratio $\lambda$.

Based on the transition rate equation, the expressions of system performance parameters such as cooling power, COP, and entropy production rate were derived analytically. The COP of the QD refrigerator reaches to the maximum $\epsilon_{max} = [\epsilon_C - \lambda(1-t)]/(2-t)$ at the exoreversible working state, while the cooling power vanishes. The $\epsilon_C$-dependent CMCP was then studied at different given $t$ and $\lambda$ numerically with $\epsilon_C \geq \lambda(1-t)$ except $t = 1$, which corresponded to the case without considering the system resistance. The small value of CMCP (less than 1) is consistent with a macroscopic thermoelectric refrigerator, though it is not a monotonic decreasing function of $\lambda$ anymore. At small $\epsilon_C$, the maximum cooling power increases as $\lambda$ decreases since less Joule heat flows into the cold reservoir, which improves CMCP distinctly. For very large $\epsilon_C$, less power input is needed to generate certain amount of maximum cooling power at larger $\lambda$, resulting in remarkable enhancement of CMCP. In other words, absorbing more Joule heat by the cold reservoir corresponds to larger CMCP, which is partially consistent with the macroscopic thermoelectric refrigerator.

This work reveals the impact of Ohmic contact resistance on the performance of a single-level QD refrigerator, especially on the COP and CMCP. The actual COP and CMCP will be smaller than the obtained results here, because other factors such as electron spin, strong coupling between QD and reservoirs, electron-electron interaction, electron-phonon interaction, other forms of dissipation etc. were not considered in our simplified QD refrigerator model. However, our results still provide useful guidance on the design and exploitation of mesoscopic refrigerators.

**Acknowledgements**

This work was supported by the National Key Research and Development Program of China (No. 2016YFA0202401), the National Natural Science Foundation of China (No. 61705066, 51772095, 51702096, U1705256), the 111 Project (No. B16016) and Fundamental Research Funds for the Central Universities (2017MS028, 2018ZD07, G2018KY0303), the National Natural Science Foundation of Shaanxi Province (No. 2019JQ-059).